# CRAQL: A Composable Language for Querying Source Code

Blake Johnson
The George Washington University
Washington, DC
bej@gwmail.gwu.edu

Rahul Simha
The George Washington University
Washington, DC
simha@gwmail.gwu.edu

## ABSTRACT

This paper describes the design and implementation of CRAQL (Composable Repository Analysis and Query Language), a new query language for source code. The growth of source code mining and its applications suggest the need for a query language that can fully utilize and correlate across the unique structure and metadata of parsed source code.

A major goal of our project was to build a language on an underlying abstraction analogous to the underpinnings of SQL, but aimed at parsed source code. Thus, while SQL queries' inputs and outputs consist of sets of tuples, CRAQL queries' inputs and outputs consist of sets of abstract syntax trees. This abstraction both makes CRAQL queries composable (the output of one query can become the input to another) and also improves the power of the language by allowing for querying of the tree structure and metadata, as well as raw source code. Furthermore, the abstraction enables tree-specific language optimizations and allows CRAQL to be easily applied to any language that is parsable into abstract syntax trees. These attributes, along with a familiar syntax and organization similar to SQL, allow complex queries to be conveniently expressed in a compact, straightforward manner. Questions such as "find the longest series of statements without any loops," "find methods that are never called," "find getters (methods with no parameters and a single statement that directly returns a member variable) in each class," or "find the percentage of variables that are declared at the top of a block" all translate into simple, easy to understand queries in CRAQL.

In this paper we describe the language, its features and capabilities. We compare CRAQL to other languages for querying source code and find that it has potential advantages in clarity and compactness. We discuss the features and optimizations we added to support searching parse tree collections more effectively and efficiently. Finally, we summarize the application of the language to millions of Java source files, the details of which are in a companion paper. We hope that this language and our associated open source implementation will prove useful to the MSR community.

## 1 INTRODUCTION

Data mining of open source code repositories is one of the fastest growing fields within software engineering [1]. However, the usefulness of these repositories is limited by the available languages for querying source code. Urma and Mycroft [2] surveyed query languages for Java and found shortcomings in most of the languages they studied – some are character based and cannot query parse trees, some do not support all structures of the desired target language, some lack the ability to query bindings and expression types, some are proprietary, and yet others require overly complex or verbose queries.

A language that allows the full richness of source code to be searched must take into account two key characteristics of source code – its syntactical structure, and its metadata and post-compilation bindings. For an example of the latter, when examining a reference to a variable, it should be possible in most languages to know that variable's type, and whether it was declared constant, and when encountering a method call, it should be possible under many circumstances to know which method definition is being called. At the same time, there are several other practical requirements for such a query language related to power, performance, and ease of use. Our main goal in this project was to design a language with the familiarity and ease of use of SQL, but operating on parse trees instead of tables, that could meet these criteria:

1) The language must offer specialized operators and optimizations to allow the user to effectively manage the tree structure of code since most programming languages compile to Abstract Syntax Trees (ASTs).
2) The language should make all derivable post-compilation metadata and bindings available for querying.
3) The language should allow fine-grained queries down to the individual syntax-token level.
4) The language must enable optimized, high performance query execution that can scale to millions or billions of source files, to support data mining of large repositories.
5) The language should be *composable* – that is, the output of a query should be able to be used as the input to another. This is a key requirement to allow complex queries to be built up from smaller, simpler ones. Furthermore, by including metadata and aggregate statistics in the parse-tree set, such information can pass through when queries are composed.
6) The language should be extensible, and allow programmers to include imperative code inline for maximum flexibility and compactness.

7) Finally, the language should have a familiar, easily-understood SQL-like syntax that will allow simple, compact construction of queries, with clear, meaningful results.

In this paper we present the Composable Repository Analysis and Query Language (CRAQL), a new, flexible query language for source code. We will discuss the design of the language, how it meets the above criteria, its optimizations, some results of testing queries on Java source code, and our open source implementation of CRAQL which may be further enhanced or integrated into other tools. We also discuss a few of the surprises we found while implementing our language, such as discovering that the tree-pruning operations we introduced to optimize query performance were the same operations we came to use constantly to tailor our results correctly in our tree-based system. This paper focuses on the language itself while a companion paper presents an analysis of the Java language based on over 100 million lines of source code, showing the language's evolution over two decades. The next section describes related work while Section 3 describes CRAQL with examples. Section 4 outlines some experimental results.

|  | CRAQL | srcQL | Boa | CQLinq | PQL | Astlog |
|---|---|---|---|---|---|---|
| Queries Tree Structure | X | X | X* |  |  | X |
| Metadata/ Bindings | X |  |  | X | X |  |
| Token Level Queries | X | X** |  |  |  | X |
| Composable | X |  |  | X |  | X |
| Supports Imperative Queries | X |  | X*** | X |  |  |
| Familiar, SQL-like | X | X |  | X | X |  |

\* Boa queries operate on a tree that is slightly abstracted from the native language tree
\*\* srcql further allows syntax pattern-matching even within a token
\*\*\* Boa queries are imperative but must be structured according to Boa's visitor paradigm.

**Figure 1: Source Code Query Languages Feature Comparison**

## 2 RELATED WORK

The querying of source code is a relatively recent phenomenon, as publicly available codebases have grown in number and size. Historically, query writers used simple text searching tools like awk and grep, or database query languages based on the relational model (like SQL), but these are inadequate for source code because they do not capture the rich structure and semantics of code. However, these sorts of techniques have nonetheless been used. For example, Google Code Search [3] offered a huge archive of code to search and regular expression queries, as well as improved handling for special characters, a step above most other options at the time, but still failed to capture the tree structure of code. Paul and Prakash [4] suggest the need for a formal model on which to base a code query language and propose their own "Source Code Algebra." Another popular way to reduce or avoid the limitations of flat searches is a natural language based search approach, where sophisticated query interpretation is correlated with the linguistic information present in source comments and identifier names. Haiduc et al [14] developed a system to automatically detect low-quality queries and rewrite them for a better chance at relevant search results. Hill [12] proposed a hybrid system combining natural language and program structure that used the natural language approach to prune likely poor results to allow a recursive exploration of the program structure to locate more promising results.

For some research purposes, not all repository mining needs to query deeply within the source code at all. In some cases, analysis of the commits to a project is sufficient to derive interesting results about project development. Robles et al [13] described how the other non-code artifacts in a project may be as rich a source of information as the code itself. In other cases only a limited reach into the structure of code is required. CQL (Code Query Language) and its successor, CQLinq, which underpin the popular NDepend static analysis tool, are SQL-like languages for .NET projects that look primarily at high level .NET assembly metadata, representing programs as simple relations [5]. They have only an extremely limited capability to search the actual code of the project below the class and member level. PQL [6] is a fascinating query language focused on the identification of object event patterns, and cannot query source code directly, except to analyze sequences of method calls and returns. Mcmillan et al [15] identify chains of function calls as the key search result developers require and their Portfolio system accordingly queries the code as a directed graph of function calls.

Query languages focused on tree or graph structures have relevance to code querying, as most source code may be parsed into ASTs, though these languages generally need additional customization to fully support querying source code. XQuery [7] is a W3C standard language based on XPath for querying XML documents, which provided a particular influence on our work due to its natural implementation of paths through the XML nodes and its straightforward and powerful FLOWR expression which allows the output to be filtered, tweaked, and output into a customized html format. PMD [8] is a source code analysis tool which is also based on XPath with many simple and useful rules predefined, and it allows efficient querying of the entire AST, but it lacks support for binding and other metadata. Gremlin [9] is a graph traversal and query language that allows a mixture of imperative and declarative queries, a flexibility that seems well suited to tree structures, that we have also drawn inspiration from.

The BOA language and infrastructure [10] is a comprehensive and highly capable system that appears unsurpassed for many source code mining use cases. It incorporates language elements specifically for repository mining, supports high performance distributed implementations, and makes streamlined ASTs available for querying. The infrastructure also provides a massive index of open source code to query. However, the visitor-based query language, derived from Google Sawzall, has a fairly steep learning curve and can result in verbose, hard-to-understand query definitions. Also, the full binding and type information is not captured, and it is not fully composable. Some similar issues affect ASTLOG [11], a Prolog variant for analysis of ASTs – the queries can be long and complicated, even for some very simple questions. Our hope is that a simpler language with a familiar syntax and yet offering full AST search and composability can offer a complementary alternative.

Another language in a similar domain to CRAQL language is the srcQL language of Bartman et al [17]. srcQL operates on srcML, an xml representation of source code. srcQL features a SQL-like, declarative query format, and not only allows searching down to the individual token level, but further allows pattern matching within the token text. This allows for extremely fine-grained queries, but compared with CRAQL, it lacks composability, exposure of bindings and metadata, and the ability to design imperative or mixed queries.

## 3 LANGUAGE DESCRIPTION

### 3.1 Overview

In CRAQL, the primary input and output data of each query are sets of Abstract Syntax Trees, along with arbitrary user defined variables. The language combines a familiar SQL-style declarative query syntax with a structured, imperative, C-like component for filtering, refinement, and pre- and post-processing of results. Both language subsets may be freely mixed in each query. Most importantly, all of the additional information that source code contains beyond regular text (hierarchies, metadata, and bindings) is available to query. The language also contains several important keywords and operators for query optimization. CRAQL is programming-language agnostic, but our initial implementation operates on Java source code. The broad structure of a CRAQL query is familiar, and modeled on SQL and the relational algebra -- selecting a subset of entities (in his case, ASTs) from a set, often winnowed by a where clause.

```
1:  // query to count block-top vs inline variable declarations
2:  // ----------------------------------------------------------------
3:  // A declaration is block-top if it is part of a variable declaration
4:  // statement, and is not preceded by any other kind of statement
5:
6:  select ({Block} b)
7:  {
8:    select ({VariableDeclaration} v) directly in b
9:    {
10:     temp_is_blocktop = 1;
11:
12:     if (v.parent().isnodetype({VariableDeclarationStatement}))
13:     {
14:       select outmost ({Statement} s) in b
15:         where (b == s.parent()) &&
16:               (s.position() < v.position()) &&
17:               !s.isnodetype({VariableDeclarationStatement}) {
18:         temp_is_blocktop = 0;
19:       }
20:     }
21:     else {
22:       temp_is_blocktop = 0;
23:     }
24:     num_blocktops += temp_is_blocktop;
25:     num_inlines += (1 - temp_is_blocktop);
26:   }
27: }
```

Annotations:
- Line 6: Select all code blocks in the
- Line 8: Select the variable declarations in each block
- Line 10: The "temp_" prefix excludes the variable from our output
- Lines 12-17: See if the results meet criteria for a block-top declaration.
- Lines 21-23: Update output variables to count the number of each kind of declaration

**Figure 2: An example CRAQL query.**

### 3.2 Guided Tour of a CRAQL Query

Figure 2 shows an example CRAQL query, designed to answer the question: how often do programmers declare variables at the top of a code block versus declaring them as needed? The overall approach is to find all blocks, find all variable declarations directly within them, and then see if any non-variable declaration statements precede them. Then we will count each type of declaration to measure its prevalence in repository code.

We first select all Blocks (or more accurately, the set of subtrees with a Block as the root node) from our source project, and all top-level Variable Declarations within them. Identifiers enclosed in curly braces (like "`{Block}`" on line 6) represent node types in the AST. The "`in`" keyword (on line 8) is used to direct the output trees from the outer query into the input for the inner query. The "`directly`" keyword, also on line 8, is a tree pruning operation, fully discussed in a later section of this paper, which eliminates those variable declarations which are enclosed within sub-blocks of each block result, as for the purposes of this query, they must be considered only with the sub-blocks they occur within, not with any outer block. This result set of Block subtrees becomes the input to the inner query for VariableDeclarations.

Each variable declaration is processed between lines 9 and 27. Here we switch briefly to (optional) imperative logic to determine whether the variable declaration is block-top or not. Line 12 uses the special CRAQL functions `parent()` and `isnodetype()` to determine if the declaration is the child of a variable declaration statement node (a prerequisite for a block-top declaration). A careful reading shows that the "`if`" statement on line 12 could be replaced by a where clause on the variable declaration query on line 8, however we chose this implementation to demonstrate the flexibility with which imperative and declarative execution may be combined. On line 14, we use the "outmost" keyword, another similar tree pruning operation further discussed below, which has the effect of only selecting top-level statements in the block, rather than nested statements (which would have no effect on the block-top property of the declaration). We also use the CRAQL "position()" function, which obtains the character position within the source file of the first character in the text comprising the AST node, to see which statement precedes the other.

### 3.2 CRAQL Language Features

```
 1: // Figure 3, snippet 1: contains() –
 2: // find catches that throw
 3: select ({CatchClause} c) {
 4:     if (c.contains({ThrowStatement})) {
...
 1: // snippet 2: contains() using a bound result tree
 2: select ({TypeDeclaration} t1) {
 3:     select ({TypeDeclaration} t2) {
 4:         if (t1.contains(t2)) {
...
 8: // snippet 3: isparent()
 9: select ({Block b}) {
10:     select ({Statement s}) in s
                    where b.isparent(s) {
...
```

**Figure 3: A CRAQL query snippet demonstrating the contains() and isparent() functions.**

Figure 3, snippet 1 contains part of a CRAQL query aimed at studying the question: how often does a try-catch actually throw an exception? This snippet shows the use of the contains() function, which returns true if a tree contains the given AST node type. This form of contains() often serves as a shortcut to avoid an additional nested query, in cases where the query writer needs to know only the existence of a particular node type within a tree.

Snippet 2 shows a second form of contains(), where the function is passed a subtree, rather than a node type, and it returns true if the parent tree contains the given subtree. The second form is only occasionally useful, as in most cases it is more efficient to flow output from one query to another using the "in" keyword, rather than determine the relationship after the fact using contains().

Snippet 3 is a query that returns statements paired with their enclosing block. The isparent() function is similar to contains(), except it only returns true when a direct child to the root matches the node type or given subtree. Like contains(), isparent() can be passed either a subtree or a node type. As with the second form of contains, isparent() can often be replaced by the use of "in", if it is combined with the more efficient "directly" and "outmost" keywords (discussed below), but as these do not exactly match the isparent() functionality, there are some cases when isparent() cannot be easily replaced in a where clause. In the future it would be interesting to see what performance and compactness improvements might follow from the addition of a new query modifier which more closely matches the behavior of isparent().

The query in Figure 4 is designed to find getters. Under our fairly strict definition, a getter is a method with no parameters and a single statement that directly returns a single variable. As it evaluates these criteria, the query shows some of the ways the query author can traverse the returned subtrees. First the query selects methods that have no parameters and a single statement. Then a second query retrieves the statements from within the methods, filtering out those that are not return statements or do not directly return a variable.

```
 1: // Figure 4: Finding getters
 2: select ({MethodDeclaration} m) where
 3:     m.parent().isnodetype({ClassDeclaration}) &&
 4:     !m.{parameters} &&
 5:     m.{body}.{statements} == 1) {
 6:   select outmost (Statements s) in m where
 7:        s.isnodetype({ReturnStatement}) &&
 8:        s.{Expression}.isnodetype({Name})) {
 9:     print(m + " is a getter");
10:   }
11: }
```

**Figure 4: CRAQL query snippet demonstrating tree traversal techniques.**

The parent() function on line 3 is used to ascend to the parent of the given node (in this case, to affirm that this method is within a class (rather than an interface). To descend the tree, as on lines 4, 5 and 8, the user may specify either the name of the subtree (or subtree list) that they wish to traverse to (as in "{parameters}", "{body}", and "{statements}"), or, as a shortcut in cases where the subnode can be unambiguously determined, by specifying the node type of the desired child node (as in "{Expression}"). In both cases these names are determined by the grammar of the target language. The check against "m.{body}.{statements} == 1" on line 5 is special because the "statements" property of a block is not an AST node, but rather a list of AST nodes. CRAQL automatically casts such a list to an integer (the number of nodes in the list) when it is compared against an integer as on line 5.

```
1: // Figure 5: Query Calls - counting loop depths
2: q1 : select outmost (ForStatement f) {
3:    nested_for_count ++;
4:    callquery(q1) directly in f;
5: ...
```

**Figure 5: A CRAQL query snippet demonstrating query calls.**

Figure 5 contains a snippet from a query designed to find the most deeply nested for loops. For this task it is necessary to recursively descend through the parse tree through all of the nested loops. Line 4 shows how the "`callquery()`" function can be used to dynamically call another query (or the same query again). In this case the query makes a recursive call to itself to count the number of nested for loops. This feature is also frequently useful in cases where a query is searching for two or more associated node types, and can be used dispatch execution to the appropriate inner query based on which type is actually encountered.

```
1: // Figure 6, snippet 1: Method bindings
2: select ({MethodDeclaration} m) {
3:     select ({MethodInvocation} i)
4:         where m == i.methodbinding() {
...
```

```
1: //  Figure 6, snippet 2:  Type bindings
2: select ({TypeDeclaration} t) {
3:     select ({Expression} e) in t
4:         where e.typebinding() == t {
...
```

**Figure 6: CRAQL query snippets demonstrating method and type bindings.**

The snippets in Figure 6 demonstrate how the method and type bindings implicit within the source code may be queried. The first snippet contains two nested queries which find method declarations paired with method calls that invoke them. The `methodbinding()` function returns the method declaration that a method invocation is calling.

Snippet 2 has a similar design, and finds type declarations paired with expressions that resolve to that type. The `typebinding()` function returns the type declaration that an expression returns. Thus, a method invocation may have both a `methodbinding()` and a `typebinding()`, which would be return type of the method being called.

Note that only bindings that may be statically determined at compile time can be queried in this manner.

```
1: // Figure 7: Counting number of in/out calls
2: select ({TypeDeclaration} t) {
3:   num_incoming = 0;
4:   mum_outgoing = 0;
5:   // find methods calls out of this class
6:   select ({MethodInvocation} m) in t
7:       where !t.isparent(m.methodbinding()) {
8:     num_outgoing++;
9:   }
10:  // now find method calls into this class
11:  select ({MethodInvocation} m2)
13:      where !t.contains(m2) &&
14:             t.isparent(m2.methodbinding()) {
15:    num_incoming++;
16:  }
```

**Figure 7: A CRAQL query snippet for measuring the number of incoming/outgoing method calls per class**

Figure 7 contains an appealingly compact query to count, for each class, the number of method calls into and out of that class' methods. It is a component of a simple, static-analysis version of the "Hubs and Authorities" webmining-based key class identification method developed by Zaidman and Demeyer [16]. The query proceeds by selecting all classes, then selects all method invocations that originate within each class but conclude outside of it. Finally it selects all method invocations that originate outside each class but conclude inside it. These operations are only possible because of the method binding information that is exposed by the `methodbinding()` function, and they demonstrate the powerful and concise queries that it enables.

### 3.2 Tree Optimizations

```
1: // Figure 8: tree pruning and optimization keywords
2: select ({MethodDeclaration} m) {
3:     select outmost ({Statement} s) directly in m {
...
```

**Figure 8: A CRAQL query snippet demonstrating tree pruning keywords.**

Figure 8 shows two of the tree-pruning keywords we added to CRAQL which may be used to tailor results and optimize search performance. For any query "`select outmost ({NodeType} n)`", the outmost keyword causes the query to only return subtrees where there are no other {NodeType} nodes interposed between the root of the input to the query and the root of the subtree being returned. The inverse of "outmost", "inmost," is also available, which has the effect of excluding subtrees where the returned subtree contains within it additional nodes of {NodeType}. An associated optimization is the "directly in" syntax, which, for any query "`select ({NodeType} n) directly in inputTree`", causes the query to return only subtrees where there are no other nodes of the type of the root node of inputTree interposed.

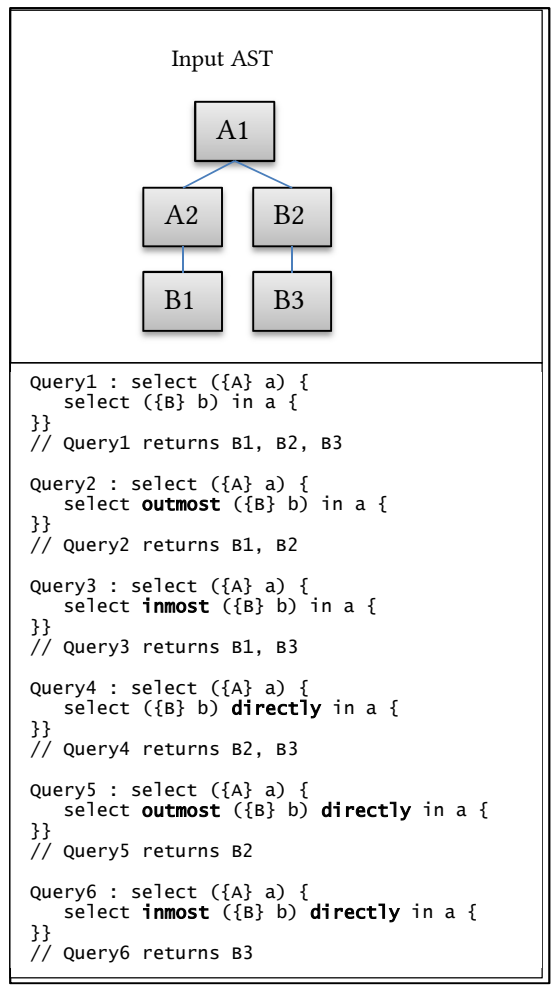

```
Query1 : select ({A} a) {
    select ({B} b) in a {
}}
// Query1 returns B1, B2, B3

Query2 : select ({A} a) {
    select outmost ({B} b) in a {
}}
// Query2 returns B1, B2

Query3 : select ({A} a) {
    select inmost ({B} b) in a {
}}
// Query3 returns B1, B3

Query4 : select ({A} a) {
    select ({B} b) directly in a {
}}
// Query4 returns B2, B3

Query5 : select ({A} a) {
    select outmost ({B} b) directly in a {
}}
// Query5 returns B2

Query6 : select ({A} a) {
    select inmost ({B} b) directly in a {
}}
// Query6 returns B3
```

**Figure 9: A comparison of the inmost, outmost, and directly keywords.**

Figure 9 illustrates the effect of these modifiers on query results. An alternative to the `contains()` function is also provided, `directly_contains()`, which yields the same effect as the directly modifier does when applied to queries.

We initially designed these modifiers to filter extraneous or duplicative subtrees and thus increase performance, and they do indeed improve performance for many queries, but they have also surprised us by being very useful functionally, as well, and often are necessary for strictly correct results. For example, the situation arises commonly (as in our initial example in Figure 2) where the query author wishes to find only the top-level blocks or statements within a method, and exclude those within nested loops or inner classes. The use of "`outmost`" and "`directly`" make this condition easy to implement.

## 4 SAMPLE IMPLEMENTATION

To explore the power and performance of CRAQL, we have built an initial open-source implementation of the parser and query processor supporting queries on Java-language source code. For the Java language parsing and metadata inference we required, we used the Eclipse Java Development Tools (JDT) libraries interfaced with our own code, also written in Java, for parsing and executing CRAQL. The system supports initializing sets of project-specific variables that will flow into the queries, may be used or modified, and finally are output along with the result subtrees and any other output variables created within the queries. After query execution, the output variables for all projects are gathered up and built into a single spreadsheet to aid in analysis.

To test the implementation, we also wrote several dozen queries to investigate areas of interest in the use of the Java language. We executed these queries on over 2 million Java source files, with well over 100 million lines of code, taken from the top 3000 Java projects on github spanning two decades of Java usage. The analysis of our query results will be presented in a companion paper, *An Archaeological Study of Java Using the CRAQL language*. In this paper we only consider the performance characteristics of this large scale execution.

Our measured performance on this dataset was reasonable for an initial, lightly optimized, single-threaded implementation, especially after the implementation of the "`inmost`", "`outmost`" and "`directly`" query modifiers, which ended up in many of our queries. The precise performance characteristics are highly dependent on the number and content of the queries and projects, as well as the hardware platform, but when executing our reasonably broad sample of about 50 queries of varying sizes on a modest PC, the system worked through about 15 million lines of code per hour. On a single simple query, representing the maximum performance of the current implementation, the system processed about 75 million lines of code per hour. The relative lack of speedup indicates that a significant portion of the total execution time in the single-query scenario was taken up by parsing, whereas parsing time is relatively unimportant in the 50 query scenario (as each project is parsed only once regardless of how many queries are executed on it).

There are many potential improvements we intend to make to this early, promising implementation. Some of the smaller convenience features of the language have yet to be implemented, and there is great scope for additional performance optimizations. It should also be reasonably straightforward to add support for C++ code by integrating the Eclipse C/C++ Development Tools (CDT) libraries, which share a common interface with the JDT libraries we currently use, and our CRAQL parsing and execution code is written to be independent of the target language (existing queries, of course, would need to be modified for the different node types in a C++ AST).

```
 1: select ({Block} b)
 2: {
 3:   select outmost ({Statement} s1) directly in b
 4:   {
 5:     select outmost ({Statement} s2) directly in b
 6:         where s1.position() < s2.position() &&
 7:        (s1.isnodetype({BreakStatement}) ||
 8:         s1.isnodetype({ReturnStatement}) ||
 9:         s1.isnodetype({ThrowStatement}) ||
10:         s1.isnodetype({ContinueStatement}))
11:     {
13:       print(s2.filename() + " - " + s2.linenumber());
15:     }
16:   }
17: }
```

**Figure 10: A CRAQL query for locating instances of unreachable code**

## 5 CONCLUSION

Our results demonstrate some of the benefits repository miners may enjoy through the use of a query language that combines the composability and familiar syntax of traditional relational query languages with an input/output model adapted for ASTs, and operations and optimizations tailored for source code, with its additional richness. Many queries that are possible in CRAQL are simply impossible in many other languages. In other cases it maybe possible to design CRAQL queries that are more compact or clear when compared against their equivalents in other languages.

For example, Figure 10 shows a query to identify certain instances unreachable code, based on an equivalent query from the reference documentation of the Boa language. The query selects all blocks, then attempts to find two statements directly within each block where the statement that precedes the other is of a type that aborts execution of the block code (specifically, the `break,` `return,` `throw,` and `continue` statements). The CRAQL query used 15 lines and 418 characters, compared with the equivalent query in the other language, which used 47 lines and 1,210 characters. Users may also find the CRAQL example to be clearer and easier to read, due to its SQL-like semantics and because it is structured and ordered the way a human would naturally perform the same search, vs the powerful but unintuitive visitor-stype query required by Boa. Our initial implementation of a CRAQL system demonstrates that these benefits are realizable without sacrificing high performance.

Further work is needed to refine our implementation and explore additional improvements to the language itself. For example, the "`directly`" and "`outmost`" keywords, which have proved immensely useful in our query set, feel like two special cases of a more general purpose operator, which would prune subtrees in which any arbitrary, user-specified AST node type (or set of node types) is interposed between the input and result tree roots. It would also be desirable to further customize these modifiers to, optionally, incorporate the behavior of our `isparent()` function. We would like to find a way to implement such a general operator without sacrificing the clarity and compactness of our query code. We are also interested in enhancing our system to operate on source code augmented with recorded trace logs of a program's execution. Combining the structure of the code with its actual real-world execution may, in many cases, enable us to query programs more effectively.

## ACKNOWLEDGMENTS

The authors would like to thank Sarah Casay and Loc Nguyen, whose senior design projects contributed to this paper.

## REFERENCES

[1] Pinzger, M. & Kim, S. Empirical Software Eng (2016) 21: 2033. https://doi.org/10.1007/s10664-016-9450-8
[2] Raoul-Gabriel Urma and Alan Mycroft. 2012. Programming language evolution via source code query languages. In Proceedings of the ACM 4th annual workshop on Evaluation and usability of programming languages and tools (PLATEAU '12). ACM, New York, NY, USA, 35-38.
[3] Cox, Russ (2012). Regular Expression Matching with a Trigram Index. Retrieved from https://swtch.com/~rsc/regexp/regexp4.html
[4] S. Paul and A. Prakash. Supporting Queries on Source Code: A Formal Framework. International Journal of Software Engineering and Knowledge Engineering, 4(3):325-348, 1994.
[5] Smacchia, Patrick A (2008). Code Query Language 1.8 Specification. Retrieved from https://www.javadepend.com/CQL.htm
[6] Michael Martin, Benjamin Livshits, and Monica S. Lam. 2005. Finding application errors and security flaws using PQL: a program query language. SIGPLAN Not. 40, 10 (October 2005), 365-383.
[7] Don Chamberlin. 2003. XQuery: a query language for XML. In Proceedings of the 2003 ACM SIGMOD international conference on Management of data (SIGMOD '03). ACM, New York, NY, USA, 682-682. DOI=http://dx.doi.org/10.1145/872757.872877
[8] PMD Introduction (2018, January 21). PMD Source Code Analyzer Project. Retrieved from https://pmd.github.io/pmd-6.0.1/
[9] Marko A. Rodriguez. 2015. The Gremlin graph traversal machine and language (invited talk). In Proceedings of the 15th Symposium on Database Programming Languages (DBPL 2015). ACM, New York, NY, USA, 1-10. DOI=http://dx.doi.org/10.1145/2815072.2815073
[10] Robert Dyer, Hoan Anh Nguyen, Hridesh Rajan, and Tien N. Nguyen. 2013. Boa: a language and infrastructure for analyzing ultra-large-scale software repositories. In Proceedings of the 2013 International Conference on Software Engineering (ICSE '13). IEEE Press, Piscataway, NJ, USA, 422-431.
[11] Roger F. Crew. 1997. ASTLOG: a language for examining abstract syntax trees. In Proceedings of the Conference on Domain-Specific Languages on Conference on Domain-Specific Languages (DSL), 1997 (DSL'97). USENIX Association, Berkeley, CA, USA, 18-18.
[12] Emily Hill. 2010. Integrating Natural Language and Program Structure Information to Improve Software Search and Exploration. Ph.D. Dissertation. University of Delaware, Newark, DE, USA.
[13] Gregorio Robles, Jesus M. Gonzalez-Barahona, and Juan Julian Merelo. 2006. Beyond source code: the importance of other artifacts in software development (a case study). *J. Syst. Softw.* 79, 9 (September 2006), 1233-1248.
[14] Sonia Haiduc, Gabriele Bavota, Andrian Marcus, Rocco Oliveto, Andrea De Lucia, and Tim Menzies. 2013. Automatic query reformulations for text retrieval in software engineering. In *Proceedings of the 2013 International Conference on Software Engineering* (ICSE '13). IEEE Press, Piscataway, NJ, USA, 842-851.
[15] Collin Mcmillan, Denys Poshyvanyk, Mark Grechanik, Qing Xie, and Chen Fu. 2013. Portfolio: Searching for relevant functions and their usages in millions of lines of code. ACM Trans. Softw. Eng.

Methodol. 22, 4, Article 37 (October 2013), 30 pages.
[16] Andy Zaidman, Serge Demeyer. Automatic Identification of Key Classes in a Software System Using Webmining Techniques.
Journal of Software Maintenance and Evolution: Research and Practice 20(6): 387-417, Wiley, November/December 2008.
[17] B. Bartman, C. D. Newman, M. L. Collard and J. I. Maletic, "srcQL: A syntax-aware query language for source code," 2017 IEEE 24th International Conference on Software Analysis, Evolution and Reengineering (SANER), Klagenfurt, 2017, pp. 467-471.

# A    Appendix: Implementation Notes

This appendix contains more detailed information about our open source implementation of CRAQL and how it may be used and enhanced by others. Our website address is temporarily withheld to align with blind submission.

## A.1    Prerequisites

CRAQL has been tested on Linux and Windows PCs. The only prerequisite software that must be installed is the Java 1.8 Runtime. Our distribution includes the required Eclipse Java Development Tools libraries, so an Eclipse installation is not required. However, we do include Eclipse project files for CRAQL, and Eclipse 4.5 or later is recommended as a development environment for testing and updating our code.

## A.2    Installation

Once the installation package has been unzipped or cloned from github, the only configuration required is the identification of a set of directories to contain project source files, query files, input properties, and results, and an updated classpath to bring in the included Eclipse JDT libraries. Figure 11 shows a typical directory hierarchy.

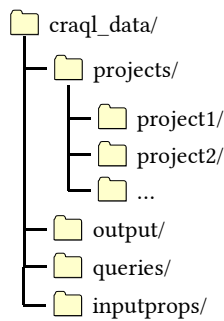

**Figure 11: A sample directory hierarchy for CRAQL input and output data**

## A.3 Execution

The current implementation lacks a way to automatically download project source code. Instead, the user must download the source code through other means and place it in the project source directory in a subdirectory with the project name.

Although the user can specify a single query and project to execute on, more commonly the user will run a list of queries over a list of projects. A command line invocation will usually be of the form:

```
>java craql -P projectslist.txt -Q querylist.txt
```

where projectslist.txt contains a list of project names and querylist.txt contains a list of query files.

Once this command line is executed, the system will execute each of the specified queries against each of the projects. A file containing each result subtree will be generated in the results directory, as well as an overall output file for each project that contains all the output variables assigned during query execution on that project. When mining repository data, these output variables are typically more valuable than the raw results. Once all query executions are complete, the user can execute our included code to collate all of the output variables for all projects into a single spreadsheet:

```
>java BuildCraqlCSV
```

## A.4 Generating Input Properties

When CRAQL begins executing on a project, it looks for a properties file with a name of the form **<projectname>.properties**. This file allows the user to tag each project with any input variables they choose. These variables will be available to their query code to read and update, and will persist through the execution of the queries, eventually ending up in the output variables for that project.

This feature allows the user to tag projects with relevant labels or other information. In our 3000 project sample, we used the input properties to mark various attributes of the projects (like whether the project was for Android, J2SE, or J2EE) that aided analysis in the output spreadsheet.

Because it quickly becomes cumbersome to create an individual property file for every project, we include a helper application to generate the properties files automatically from an input spreadsheet, where the columns are the properties, and the rows are the projects.

```
>java BuildCraqlProjectProps
```

In the current implementation, the csv must be in the input properties directory with the name of projecttags.csv.

## A.5 Modifying the Implementation to Support Different Langauges

Although CRAQL itself is language agnostic, our current implementation is tied to the Eclipse Java Development Tools. Our reliance on Eclipse makes support for other languages dependent on the languages supported by Eclipse. The most obvious languages to add support for first would be C and C++, which are handled by the Eclipse C/C++ Development Tools (CDT), a library with mature and advanced support for parsing C++ as well as identifying associated type and method bindings. The CDT shares a common interface with the JDT and so should be reasonably simple to integrate with CRAQL.

The first step would be to replace package references to the JDT with equivalent references to the CDT throughout the CRAQL project. References to the JDT ASTNode type should be replaced with the CDT's IASTNode. Most of these references are encapsulated within our ASTNodeOrNodeList class, limiting the scope of this change. Finally, the parsing code in the GenerateInitialTree() method would need to be rewritten to use the CDT's ITranslationUnit parser rather than the JDT's ASTParser.

At this point, we expect a small number of compiler errors would appear where we have inadvertently relied on a JDT specific feature. After these issues are examined and resolved, the system will be ready to test with C++ source code.

It is important to note that CRAQL queries intended for Java must be refactored to reflect the different node types and structure of C/C++.

## B  Appendix: Additional Language Features

This appendix contains information about some of the more specialized features of the CRAQL language.

### B.1  Special Tree Operators

The CRAQL language contains two special query operators, the star ("*") and ellipsis ("…") operators, for more precise subtree selections. These operators are used to select subtrees with a root and a specific descendant with specific user-defined node types, and as a result, two variables are bound in each result tree.

```
1: // Figure 10, snippet 1: "*" operator – identify
2: // instances of recursion within a single method
3: select ({MethodDeclaration} decl *
            {MethodInvocation} call)
     where call.methodbinding() == decl {
...
```

```
1: // Figure 10, snippet 2: "..." operator - Find
2: // deepest nested Block
3: select ({MethodDeclaration} m ... {Block} b) {
4:   block_depth = b.depth() - m.{Block}.depth();
5:   deepest_block_depth = max(block_depth,
6:                             deepest_block_depth);
7: }
```

**Figure 12: Query snippets showing the use of the star and ellipsis query operators.**

Figure 12 demonstrates the use of both of these operators. Snippet 1 uses the star operator to identify single-method recursion (methods that call themselves). To identify single-method recursion, we must identify a pair of nodes: a method declaration and a method invocation within it that calls the parent method. The star operator simplifies finding these paired nodes. When the star operator is used, as in:

```
select ({NodeType1} n1 * {NodeType2} n2) {
```

the query will return a pair of result trees (n1 and n2) for every pair of NodeType1 and NodeType2 where the instance of NodeType2 is contained within a subtree with a root of NodeType1. The star operator is often a cleaner and more efficient substitute for the composition of queries.

The ellipsis operator is similar to the star operator, but more specialized. Figure 12, Snippet 2, uses the ellipsis operator to find the most deeply nested block. Here we need to find a method declaration, and then the deepest block within its subtree. The ellipsis operator is well suited for this task. When the ellipses operator is used, as in:

```
select ({NodeType1} n1 ... {NodeType2} n2) {
```

the query will return only the pair(s) of matching subtrees (with n2 contained within n1's tree) with the maximum distance, or depth, between n1 and n2. In snippet 2, the operator will return the method and block with the greatest depth between them, and a simple application of the depth() function is sufficient to tell us the greatest block depth. This operator is significantly more efficient than even the star operator, because it filters both the inner and outer nodes, and is frequently useful in cases where only a longest branch is required (when finding deeply nested loops or blocks, etc.).

One avenue to make the ellipses operator even more useful would be a new variant of the ellipses which would count the number of interposing instances of a specified node type, rather than absolute tree depth.

### B.1  Additional Functions

Section 3 covered many of the most common functions of the CRAQL language, including the contains(), directly_contains(), isparent(), methodbinding(), typebinding(), callquery(), and position() functions. Here we list the other functions that are available to use in queries in both declarative and imperative modes.

```
1: // Figure 12: using count(*) to limit results to 100
subtrees
2: select ({Block} b) where count(*) < 100
3: {
4:   print (count(*)); // will be 100 or less
5: }
```

**Figure 13: An interesting way to use count(*) to limit the number of search results**

**depth(**node**)** – returns the overall depth from the root at the given node.

**count(\*)** – returns the number of subtrees obtained by the current query. As with all CRAQL functions, count(*) is available in declarative query where clauses as well as imperative query code. For this reason, in addition to its obvious uses, the function provides an interesting way to limit the overall number of results, as shown in Figure 13. This technique is possible because count(*) dynamically updates as result trees are collected.

**nodetype(**node**)** – returns the AST Node type of the given node (useful for output or comparing the types of two different nodes within a query)

**max**(expression, expression), **min**(expression, expression) – Mathematical max and min functions. Note that expression may be a node if that node resolves to a numeric literal.

node.**linenumber()** – similar to position(), except it returns the line number, rather than the character position, of the first character of the text comprising the AST node.

node.**filename()** – returns a string with the file name that node was parsed from.

**print**(expression) – outputs a string or expression to the console. If expression is an AST node, prints the text from the source code comprising the node. Expressions may be concatenated with "+".

## B.2 Imperative Querying with Loops

```
1: // Figure 14: Imperatively find longest call chain
2: select outmost ({MethodInvocation} m) {
3:   temp_chain_length = 0;
4:   while (m.isnodetype({MethodInvocation})) {
5:     temp_chain_length ++;
6:     max_chain_length = max(max_chain_length,
7:                            temp_chain_length);
8:     m = m.{expression};
9:   }
10: }
```

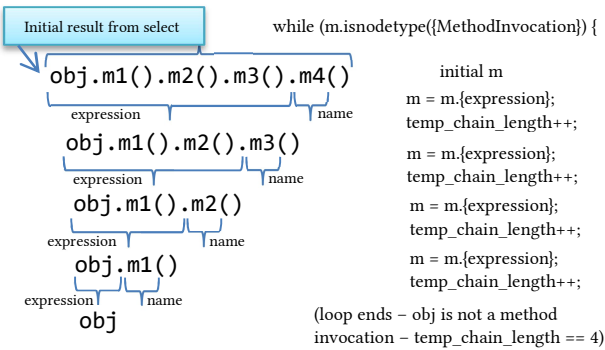

**Figure 14: A CRAQL query demonstrating iterative inner query execution using a while loop to find long method call chains.**

Generally, the clearest and most compact CRAQL queries execute primarily using our declarative syntax, with imperative code included mainly to filter and process results. However, it is possible to direct query execution manually with imperative code, often using a loop as shown in Figure 13, which shows a query to find the longest method call chain (a sequence of method calls of the form "a.m1().m2().m3()…".

This query could be expressed in declarative form using composed, recursively executed select queries (utilizing the callquery() function described in section 3.2). However in this case we chose to manually descend through the AST to locate the desired nodes using a while loop. The query in Figure 13 starts by selecting top level method invocations, and then iteratively descends the parse tree via the method invocation's expression subnode to count how many method invocations appear in a row.

Loops like this can also be used to iterate up the parse tree using the parent() function, which is useful when comparing certain kinds of depths between related nodes, or performing custom traversals where the node types may not be known in advance. CRAQL even allows queries to ascend the tree above the subtree root for each result.

Queries written with an imperative querying technique are generally more difficult to read, and more error-prone, than declarative queries, but they do usually offer more efficient execution.